\documentclass[lettersize,journal]{IEEEtran}

\usepackage{csquotes}
\usepackage{amsmath}
\usepackage{amsmath, amssymb, amsthm, bm}
\usepackage{csquotes}
\usepackage{amsmath}

\usepackage{authblk}
\usepackage{hyperref}
\usepackage{float}
\usepackage{cite}
\usepackage{algorithm}
\usepackage{algorithmicx, algpseudocode}
\usepackage{multirow}

\usepackage{mathtools}
\ifCLASSINFOpdf
\usepackage[pdftex]{graphicx}
\else
\fi
\usepackage{amsmath}
\usepackage{amsfonts}
\usepackage{amssymb}
\usepackage{color}
\usepackage{array}
\usepackage{fixltx2e}
\usepackage{stfloats}
\usepackage{url}
\usepackage{amsmath, amsthm, amssymb}

\usepackage{booktabs}
\usepackage{varwidth}
\usepackage[table,xcdraw]{xcolor}

\usepackage{color,soul}
\usepackage{caption}
\usepackage{subcaption}
\usepackage[font=scriptsize]{caption}
\usepackage{pifont}

\usepackage{adjustbox}
\usepackage{tabularx}

\begin{document}

\title{Sparsity Realization in User-Side Multilayer RIS}


\author{Hasan M. Boudi,~\IEEEmembership{Student Member,~IEEE}, and Taissir Y. Elganimi,~\IEEEmembership{Senior Member,~IEEE}

\thanks{The authors are with the Department of Electrical and Electronic Engineering, University of Tripoli, Libya (e-mail: \{h.bodi and t.elganimi\}@uot.edu.ly).}
}




\maketitle

\begin{abstract}

User-side reconfigurable intelligent surface (US-RIS)-aided communication has recently emerged as a promising solution to overcome the high hardware cost and physical size limitations of large-scale user side antenna arrays. This letter proposes, for the first time, a framework that realizes sparsity in multilayer US-RIS using two strategies, namely element-wise sparsity and geometric sparsity. The element-wise approach distributes a limited number of active elements irregularly across multiple layers, thereby exploiting additional spatial degrees of freedom and boosting the achievable rate. For further performance enhancement, a novel foldable RIS architecture leveraging geometric sparsity is proposed, achieving additional gains by optimizing the folding topology of its multilayer structure. Simulation results show that the proposed sparse architectures provide consistently higher achievable rates than existing designs.

\end{abstract}


\section{Introduction}

\IEEEPARstart{T}{he} ever-growing demand for higher data rates and more reliable wireless connections is being addressed through the deployment of large-scale antenna arrays, typically at the base station (BS). These arrays provide considerable gains in both spectral efficiency and energy efficiency \cite{massive}. However, employing large-scale arrays at the user side has been considered impractical due to spatial and power constraints. To circumvent the dimensional limitations of employing large-scale antenna arrays directly at the user equipment, the concept of a user-side reconfigurable intelligent surface (US-RIS) is proposed in \cite{9685418, User-specific}. This approach employs a transmissive RIS at the user terminal with a multilayer structure for compact implementation, thereby creating a compact high performance user device and enhancing uplink transmission capability. Significant research attention has recently been directed toward US-RIS-aided communications, motivated by their potential to circumvent the prohibitive hardware cost and physical constraints of conventional large-scale user-side arrays \cite{10196322, ES-RIS, 10930393}.

Existing advancements in the BS antenna array architectures demonstrate that substantial performance gains can be achieved not only by increasing the number of antenna elements but also through intelligent geometric arrangement \cite{10930389}. The concept of sparse arrays demonstrates that by removing the conventional half-wavelength spacing constraint, an array can achieve a much larger physical aperture with the same number of elements, yielding a reduction in inter-user interference \cite{10930389}. On the other hand, an irregular RIS architecture is proposed in \cite{irregular}, inspired by the irregular configuration of phased arrays \cite{8061016}, to overcome the constraints of regular RIS designs. The key innovation of this approach is the nonuniform distribution of the passive elements over an enlarged surface, which offers superior spatial design freedom by implementing the sparse deployment of RIS elements. However, realizing the benefits of sparsity in compact user equipment remains challenging due to spatial constraints.

In this letter, motivated by the advantages of US-RIS-aided communication with a multilayer structure \cite{10930393}, and inspired by the potential of irregular RIS architectures to enhance signal quality and suppress interference \cite{irregular}, we propose two distinct forms of sparsity for the US-RIS, namely element-wise sparsity and geometric sparsity. The element-wise sparsity is inspired by irregular RIS layouts, while the geometric sparsity is enabled by introducing a novel physically foldable multilayer architecture. To the best of our knowledge, the foldable sparse US-RIS has not been investigated in the literature; it is studied in this paper for the first time. The proposed foldable US-RIS architecture is particularly advantageous for larger user equipment categories, such as vehicular communication modules and customer premises equipment (CPE) for fixed wireless access. These platforms possess the physical dimensions required to support the proposed low-power mechanical actuators. This letter develops the system models and optimization problems for the two distinct sparse architectures. Specifically, we begin with the element-wise sparse framework and introduce sparse US-RIS. Subsequently, we extend the sparse US-RIS to incorporate geometric sparsity through a novel foldable design, which is referred to as the foldable sparse US-RIS. The integration of these two sparsity paradigms results in a joint optimization problem. Simulation results verify that the proposed sparse architectures achieve superior performance compared to existing designs.

\section{System Model And Problem Formulation} \label{sec2}

To realize sparsity in US-RIS-aided communications with multilayer structures, we first introduce the concepts of the sparse US-RIS and foldable sparse US-RIS architectures and analyze the corresponding system model. We then formulate the achievable rate maximization problem. We consider an uplink communication scenario in which the user is equipped with $K$ antennas and the BS has $M$ antennas, aided by a multilayer US-RIS. The US-RIS consists of $L$ layers, each with $N_{G}$ available grid locations. Our proposed model incorporates two distinct forms of sparsity to enhance performance while maintaining a compact design, as described below.

\subsection{Sparse US-RIS Model}

\begin{figure*}[t]
\centering
\subfloat[\label{figure_1a}]
{\includegraphics[width=7.8cm,height=5.1cm]{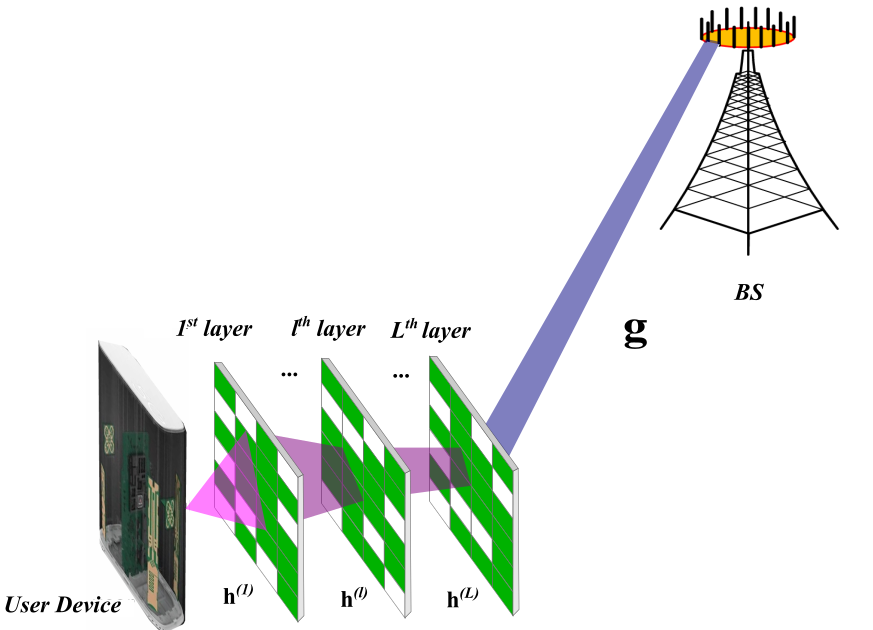}}
\subfloat[\label{figure_1b}]
{\includegraphics[width=7.8cm,height=5.1cm]{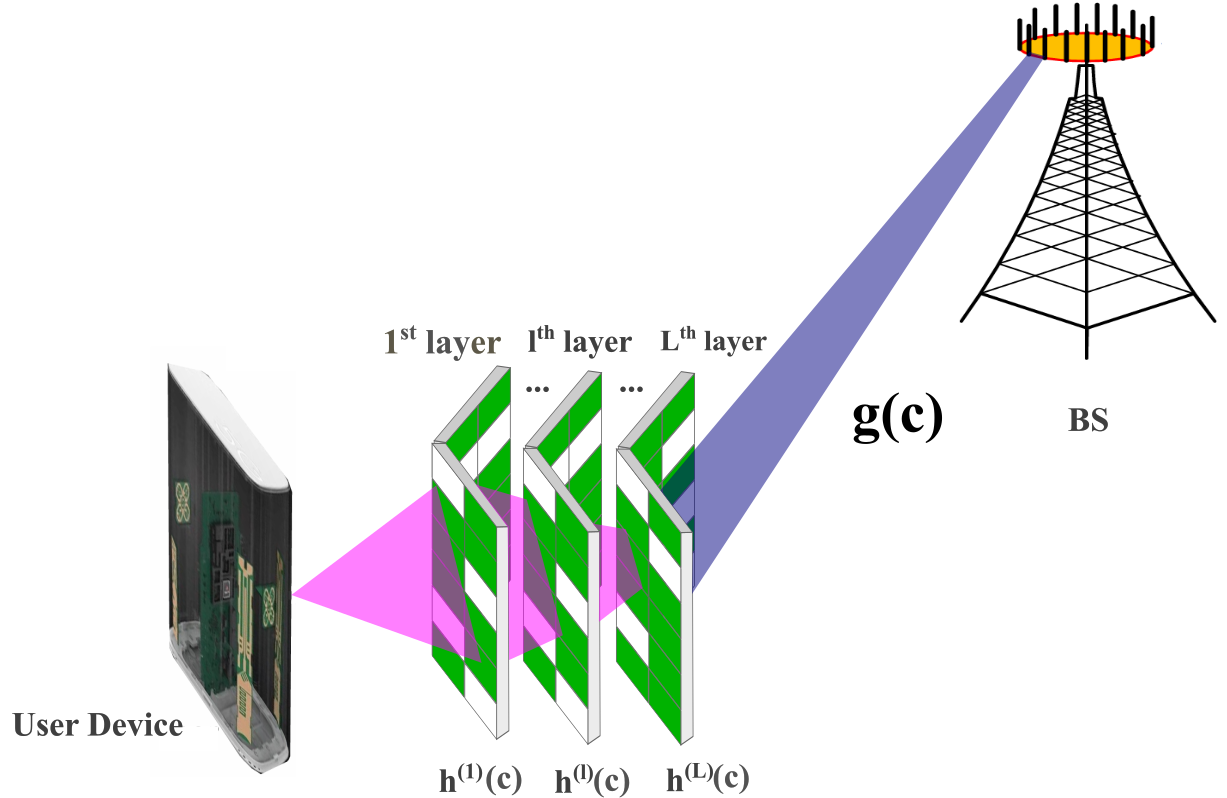}}
\caption{Illustration of (a) sparse US-RIS-aided communications; and (b) foldable sparse US-RIS-aided communications, showing the user device, the multilayer RIS, and the BS.}
\label{fig:element_wise_model}
\end{figure*}

First, we introduce the element-wise sparsity by adapting the irregular RIS framework \cite{irregular} to a multilayer US-RIS as depicted in Fig. \ref{figure_1a}. The fundamental idea is to distribute a fixed budget of $N_A$ active elements across $L$ layers, each offering $N_G$ available grid locations. This architecture, which we call the sparse US-RIS, gains additional degrees of freedom (DoFs) by strategically selecting the optimal topology of these $N_A$ elements. This, in turn, enhances the capacity compared to the conventional dense US-RIS architecture using the same number of active elements.

To formally model this architecture, we define a global activation vector, $\mathbf{z}$, which is the concatenation of the activation sub-vectors for each layer as follows
\begin{equation}
\mathbf{z} = [\mathbf{z}^{(1)T}, \mathbf{z}^{(2)T}, \dots, \mathbf{z}^{(L)T}]^T,
\end{equation}
where $\mathbf{z}^{(l)} = [z_1^{(l)}, z_2^{(l)}, \dots, z_N^{(l)}]^T$ is the sub-vector for the $l$-th layer, and $z_n^{(l)} \in \{0, 1\}$ indicates whether the $n$-th element on the $l$-th layer is activated ($z_n^{(l)} = 1$) or deactivated ($z_n^{(l)} = 0$). Based on this, we define the topology matrix for the $l$-th layer as $\mathbf{Z}^{(l)} = \text{diag}(\mathbf{z}^{(l)})$, which acts as a selection matrix.

\subsection{Foldable Sparse US-RIS Model}
\begin{figure}[t]
  \centering
  \includegraphics[width=7.3cm,height=3.8cm]{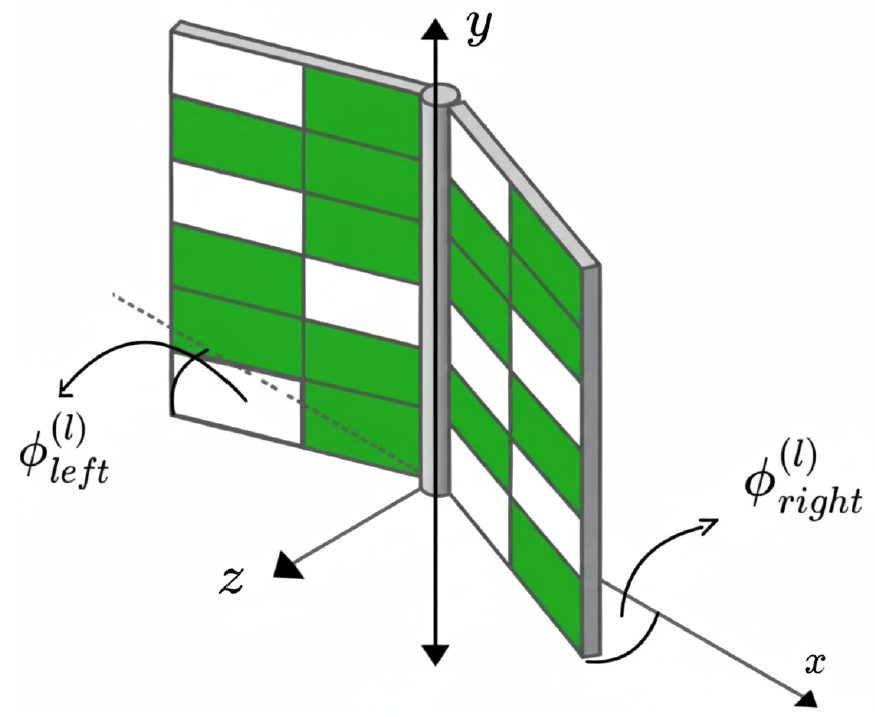}
  \caption{Illustration of the foldable sparse RIS architecture for the $l$-th layer.}
  \label{fig:foldable_layer}
\end{figure}

Building upon the sparse US-RIS framework, we introduce an additional DoF by transforming the inter-layer spacing into an optimizable parameter. This is achieved by incorporating geometric sparsity through a physically foldable architecture. The communication scenario for this architecture is shown in Fig. \ref{figure_1b}. The resultant design, which we term the foldable sparse US-RIS, simultaneously exploits both element-wise and geometric sparsity to enhance system performance. In this architecture, each layer is divided into two independently tiltable halves, namely the left half and the right half, which can be folded via a central mechanical hinge, as shown in Fig.~\ref{fig:foldable_layer}. The degree of folding for each half of each layer is defined by a specific angle. To avoid the effect of mutual coupling, we constrain the geometry by two layer-independent folding angles, $\phi_\text{left}$ and $\phi_\text{right}$. These represent the uniform folding angles applied to the left and right halves of each layer in the structure, respectively. These angles together determine the complete spatial orientation, which is captured by the configuration vector $\mathbf{c} = [\phi_{\text{left}}, \; \phi_{\text{right}}]$. They are chosen from a feasible, discrete set of $m$ possible angles, $\mathcal{A} = \{\phi_j\}_{j=1}^{m}$. 

Under the assumption of uniform spacing between the user and the first layer and between all subsequent layers, the discrete angles can be calculated as
\begin{equation}
\phi_j = -\phi_{\text{max}} + \frac{2(j-1)\phi_{\text{max}}}{m-1},
\end{equation}
where $D$ is the inter-layer spacing, and $\phi_{\text{max}}$ represents the maximum possible fold angle. This geometric constraint determines the maximum tilt angle as
\begin{equation} 
\phi_{\text{max}} = \tan^{-1}\left(\frac{D}{W_{\text{RIS}}/2}\right) = \tan^{-1}\left(\frac{2D}{W_{\text{RIS}}}\right),
\end{equation}
where $W_{\text{RIS}}$ denotes the total width of a layer in its flat case. 

In this system model, the transmitted symbol $s$ is first precoded by the beamforming vector $\mathbf{w} \in \mathbb{C}^{K \times 1}$ with power constraint $\|\mathbf{w}\|_{2}^{2} \leq P_{\max}$. The signal then propagates through the $L$ layers of the US-RIS, 
where its phase is manipulated by the phase-shifting matrix $\mathbf{\Theta}^{(l)} = \text{diag}([\theta_1^{(l)}, \dots, \theta_N^{(l)}]^T)$ at each layer, where $\theta_n^{(l)}$ is the phase shift of the $n$-th element on the $l$-th layer. The received signal at the BS is thus a function of both the element selection and the physical geometry as follows
\begin{equation}
\mathbf{y} = \mathbf{g}^H(\mathbf{c}) \left( \prod_{l=L}^{1} \alpha \mathbf{Z}^{(l)} \mathbf{\Theta}^{(l)} \mathbf{h}^{(l)}(\mathbf{c}) \right) \mathbf{w}s + \mathbf{n},
\end{equation}
where $\alpha$ is the penetration loss factor, and $\mathbf{n}$ is the additive white Gaussian noise. In this model, the channel from the user to the first RIS layer $\mathbf{h}^{(1)}(\mathbf{c}) \in \mathbb{C}^{N \times K}$, the inter-layer channels, $\mathbf{h}^{(l)}(\mathbf{c}) \in \mathbb{C}^{N \times N}$, and the channel from the last RIS layer to the BS, $\mathbf{g}(\mathbf{c}) \in \mathbb{C}^{N \times M}$, are all dependent on the configuration vector $\mathbf{c}$. The final combined signal at the BS, after applying the receiver combining vector $\mathbf{v} \in \mathbb{C}^{M \times 1}$, is modeled as
\begin{equation}   
r = \mathbf{v}^H \mathbf{g}^H(\mathbf{c}) \left( \prod_{l=L}^{1} \alpha \mathbf{Z}^{(l)} \mathbf{\Theta}^{(l)} \mathbf{h}^{(l)}(\mathbf{c}) \right) \mathbf{w}s + \mathbf{v}^H \mathbf{n}.
\end{equation}

\subsection{Achievable Rate Maximization Problem}
Our objective is to maximize the achievable uplink rate of the user. Since the achievable rate is a monotonically increasing function of the signal-to-noise ratio (SNR), maximizing the rate is equivalent to maximizing the SNR at the BS receiver. Based on the system model derived in the previous subsection, the decoding SNR can be expressed as
\begin{equation}
\text{SNR} = \frac{\left| \mathbf{v}^H \mathbf{g}^H(\mathbf{c}) \left( \prod_{l=L}^{1} \alpha \mathbf{Z}^{(l)} \mathbf{\Theta}^{(l)} \mathbf{h}^{(l)}(\mathbf{c}) \right) \mathbf{w} \right|^2}{||\mathbf{v}^H||_2^2 \; \sigma^2},
\label{eq:snr}
\end{equation}
where $\sigma^2$ represents the noise power.

The optimization problem involves finding the optimal configuration across multiple variables: $\mathbf{w}$, $\mathbf{\Theta}^{(l)}$, $\mathbf{v}$, $\mathbf{z}$, and $\mathbf{c}$. This leads to the following joint optimization problem
\setcounter{equation}{7}
\begin{IEEEeqnarray}{rCl}
\mathcal{P}: \max_{\mathbf{v}, \mathbf{\Theta}^{(l)}, \mathbf{w}, \mathbf{z}^{(l)}, \mathbf{c}} && \quad \text{SNR} \IEEEyessubnumber \label{eq:opt_problem_obj} \\
\text{s.t.} \quad && C_1: \|\mathbf{w}\|_{2}^{2} \leq P_{\max}, \IEEEyessubnumber \label{eq:constraint_power} \\
&& C_2: |\theta_n^{(l)}| = 1, \quad \forall l, n, \IEEEyessubnumber \label{eq:constraint_phase} \\
&& C_3: z_n^{(l)} \in \{0, 1\}, \quad \forall l, n, \IEEEyessubnumber \label{eq:constraint_binary} \\
&& C_4: \sum_{l=1}^{L} \sum_{n=1}^{N} z_n^{(l)} = N_A, \IEEEyessubnumber \label{eq:constraint_budget} \\
&& C_5: \mathbf{c} = [\phi_{\text{left}}, \phi_{\text{right}}], \IEEEyessubnumber \label{eq:constraint_fold_structure} \\
&& C_6: \phi_{\text{left}}, \phi_{\text{right}} \in \mathcal{A}. \IEEEyessubnumber \label{eq:constraint_fold_angles}
\end{IEEEeqnarray}

These constraints ensure physical feasibility: $C_1$ limits the user's transmit power, $C_2$ enforces the unit-modulus property of the RIS elements, $C_3$ dictates the binary nature of element activation, $C_4$ imposes the fixed budget $N_A$ for the total number of active elements, $C_5$ defines the structure of the folding configuration, and $C_6$ restricts the folding angles to a predefined discrete set $\mathcal{A}$. The optimization problem in (7) is highly complex and non-convex. In the subsequent section, we propose a joint optimization framework to obtain a sub-optimal solution. To ensure practical feasibility, we assume a two-timescale framework. The mechanical folding configuration, $\mathbf{c}$, is adapted on a slow timescale, effectively aligning the RIS aperture with long-term angular statistics, while the phase-shift parameters, $\mathbf{\Theta}$, are updated on a fast timescale to compensate for instantaneous changes. This separation aims to leverage the benefits of folding with minimal additional power consumption.

\section{Proposed Joint Optimization Framework} \label{sec3}

In this section, we propose a sequential optimization framework to decouple and solve problem (7).

\subsection{Stage 1: Element-Wise Topology Optimization}

This stage optimizes the element activation vector, $\mathbf{z}$, for a fixed initial folding topology, $\mathbf{c}_{\text{init}}$ (e.g., flat). Inspired by \cite{irregular}, we employ the tabu search algorithm iteratively as follows

\begin{enumerate}

\item  \textbf{Initialization}: We first generate a random initial feasible activation vector, $\mathbf{z}_0$, with $N_A$ active elements. Then, we calculate its achievable rate using the alternating optimization (AO) beamforming (as detailed in Subsection III-C). After that, we set this as the initial best-so-far solution, $\mathbf{z}^*$, and initialize an empty tabu list.

\item \textbf{Neighborhood Generation and Evaluation}: In each iteration, the algorithm explores the "neighborhood" of the current solution, $\mathbf{z}$. A neighbor is generated by swapping $d$ active elements (ones) with $d$ inactive elements (zeros) within the vector $\mathbf{z}$, where the parameter $d$ is the neighbor distance. A set of $S$ neighbors is generated in this manner, and for each neighbor, the algorithm first checks if it is in the tabu list.

\item \textbf{Tabu List and Candidate Selection}: Any generated neighbor found in the tabu list is discarded. For each of the remaining valid neighbors, the achievable rate is calculated using the AO algorithm. Then the neighbor that yields the highest rate among the valid neighbors is selected, which is denoted as $\mathbf{z}_{\text{next}}$ for the next iteration. After a new solution is selected, the tabu list is updated, and the previous solution is added to the list. If the list reaches its maximum size, the oldest entry is removed. 

\item \textbf{Termination}: The iterative search process continues until a predefined termination criterion is met. The algorithm then outputs the best topology, $\mathbf{z}^*$, found throughout the entire search.
\end{enumerate}
    
The output is the optimized sparse topology, $\mathbf{z}^*$, for the given $\mathbf{c}_{\text{init}}$, which serves as an input for Stage 2.

\subsection{Stage 2: Geometric Folding Optimization}

With the optimized element topology, $\mathbf{z}^*$, obtained from Stage 1, this stage optimizes the folding configuration, $\mathbf{c}$, to maximize the achievable rate as follows
\begin{enumerate}
    \item \textbf{Initialization:} Starting with an initial folding topology $\mathbf{c}_{\text{init}}$, we can calculate its rate using the AO beamforming (Subsection III-C). After that, we set this as the best-so-far $\mathbf{c}^*$, and then initialize an empty tabu list.

    \item \textbf{Neighborhood Generation:} Generating $S'$ neighbors by performing a "move" on the current configuration $\mathbf{c}$. This can be obtained by changing one folding angle, $\phi_{\text{left/right}}$, to another value in $\mathcal{A}$.

    \item \textbf{Candidate Evaluation and Selection:} For each neighbor not in the tabu list, we calculate the achievable rate using the AO algorithm. Then, selecting the valid neighbor $\mathbf{c}_{\text{next}}$ with the highest rate.

    \item \textbf{Update:} By setting $\mathbf{c} = \mathbf{c}_{\text{next}}$, if $\mathbf{c}$ outperforms $\mathbf{c}^*$, we update $\mathbf{c}^* = \mathbf{c}$, and then update the tabu list.

    \item \textbf{Termination:} We repeat Steps 2-4 until the maximum iteration count is reached.
\end{enumerate}

The output is the optimized folding topology, $\mathbf{c}^*$, for the given $\mathbf{z}^*$.

\subsection{Beamforming via Alternating Optimization (AO)}
For a fixed element topology, $\mathbf{z}$, and folding configuration, $\mathbf{c}$, the sub-problem of optimizing the beamforming variables ($\mathbf{v}$, $\mathbf{\Theta}$, $\mathbf{w}$) is solved using AO. This approach iteratively optimizes each variable while holding the others constant until convergence. The optimal solutions for each step are derived below. For notational convenience, let's first define the cascaded transmission matrix from layer $q$ to layer $p$ as 
\begin{equation} 
\mathbf{T}_{(p,q)} = 
\begin{cases} 
\prod_{l=p}^{q} \alpha \mathbf{Z}^{(l)} \mathbf{\Theta}^{(l)} \mathbf{h}^{(l)}(\mathbf{c}), & p, q \in [L], \\ \mathbf{I}_{N}, & p=L, \quad q=L+1, \\ \mathbf{I}_{K}, & p=0, \quad q=1.
\end{cases} 
\end{equation} 

\subsubsection{Optimal Receiver Combining Vector ($\mathbf{v}^*$)} 
With all other variables fixed, the SNR maximization problem with respect to $\mathbf{v}$ can be written as
\begin{multline}
\max_{\mathbf{v}} \frac{|\mathbf{v}^H \mathbf{g}^H(\mathbf{c}) \mathbf{T}_{(L,1)} \mathbf{w}|^2}{||\mathbf{v}^H||_2^2 \; \sigma^2} \\
= \frac{\mathbf{v}^H \big(\mathbf{g}^H(\mathbf{c}) \mathbf{T}_{(L,1)} \mathbf{w} \mathbf{w}^H \mathbf{T}_{(L,1)}^H \mathbf{g}(\mathbf{c}) \big) \mathbf{v}}{||\mathbf{v}^H||_2^2 \; \sigma^2}.
\end{multline}
This is a standard generalized Rayleigh quotient problem. The optimal solution is achieved when $\mathbf{v}$ is the principal eigenvector of the matrix $\mathbf{g}^H(\mathbf{c}) \mathbf{T}_{(L,1)} \mathbf{w} \mathbf{w}^H \mathbf{T}_{(L,1)}^H \mathbf{g}(\mathbf{c})$. Thus, the optimal receiver combining vector is \begin{equation} 
\mathbf{v}^* = \psi_{\text{max}}\big(\mathbf{g}^H(\mathbf{c}) \mathbf{T}_{(L,1)} \mathbf{w} \mathbf{w}^H \mathbf{T}_{(L,1)}^H \mathbf{g}(\mathbf{c}) \big), 
\end{equation}
where $\psi_{\text{max}}(\mathbf{A})$ is the eigenvector of matrix $\mathbf{A}$ corresponding to its largest eigenvalue.

\subsubsection{Optimal Phase Shifts ($\mathbf{\theta}^{(l)*}$)} 
To find the optimal phase shifts for a specific layer $l$, we rewrite the cascaded channel to isolate $\mathbf{\theta}^{(l)}$. Using the identity $\text{diag}(\mathbf{y})\mathbf{x} = \text{diag}(\mathbf{x})\mathbf{y}$ and noting that $\mathbf{Z}^{(l)}\mathbf{\Theta}^{(l)}\mathbf{x} = \text{diag}(\mathbf{z}^{(l)} \odot \mathbf{x})\mathbf{\theta}^{(l)}$, the SNR expression becomes 
\begin{multline}
\text{SNR} =
\\
\frac{|\mathbf{v}^H \mathbf{g}^H(\mathbf{c})\mathbf{T}_{(L, l+1)} \text{diag}\big(\mathbf{z}^{(l)} \odot (\mathbf{h}^{(l)}(\mathbf{c}) \mathbf{T}_{(l-1,1)} \mathbf{w}) \big) \mathbf{\theta}^{(l)}|^2}{||\mathbf{v}^H||_2^2 \; \sigma^2},
\end{multline}
where $\odot$ is the Hadamard product.

To maximize this expression, we need to maximize the magnitude of the complex scalar in the numerator. This is achieved by aligning the phases of $\mathbf{\theta}^{(l)}$ to be the conjugate of the vector it is multiplied by. Therefore, the optimal phase shift vector for layer $l$ is
\begin{equation} \mathbf{\theta}^{(l)*} = e^{j \arg \big(\text{diag} \big(\mathbf{z}^{(l)} \odot (\mathbf{h}^{(l)}(\mathbf{c}) \mathbf{T}_{(l-1,1)} \mathbf{w}) \big) \big)^H \mathbf{T}_{(L, l+1)}^H \mathbf{g}(\mathbf{c}) \mathbf{v}}. 
\end{equation} 

\subsubsection{Optimal Transmit Beamforming Vector ($\mathbf{w}^*$)} 
With $\mathbf{v}$ and all $\mathbf{\Theta}^{(l)}$ fixed, the optimization problem for $\mathbf{w}$ is 
\begin{subequations}
\begin{align}
\max_{\mathbf{w}} \quad & \frac{|\mathbf{v}^H \mathbf{g}^H(\mathbf{c}) \mathbf{T}_{(L,1)} \mathbf{w}|^2}{||\mathbf{v}^H||_2^2 \; \sigma^2} \\
\text{s.t.} \quad & ||\mathbf{w}||_2^2 \leq P_{\text{max}}.
\end{align}
\end{subequations}
 
Letting $\mathbf{a}^H = \mathbf{v}^H \mathbf{g}^H(\mathbf{c}) \mathbf{T}_{(L,1)}$, the objective is to maximize $|\mathbf{a}^H \mathbf{w}|$ by the Cauchy-Schwarz inequality, $|\mathbf{a}^H \mathbf{w}| \leq ||\mathbf{a}^H||_2 ||\mathbf{w}||_2$. This maximum is achieved when $\mathbf{w}$ is collinear with $\mathbf{a}$. To satisfy the power constraint with equality, the optimal transmit beamformer can be found as
\begin{equation} \mathbf{w}^* = \sqrt{P_{\text{max}}} \cdot \frac{\mathbf{a}}{||\mathbf{a}||_2} = \sqrt{P_{\text{max}}} \cdot \frac{\mathbf{T}_{(L,1)}^H \mathbf{g}(\mathbf{c}) \mathbf{v}}{||\mathbf{T}_{(L,1)}^H \mathbf{g} \mathbf{v}||_2}. 
\end{equation}

The AO algorithm follows these three steps to iteratively update $\mathbf{v}$, all $\mathbf{\theta}^{(l)}$, and $\mathbf{w}$, until the SNR converges.

\begin{figure}[t]
\centering
\includegraphics[width=7.8cm,height=5.3cm]{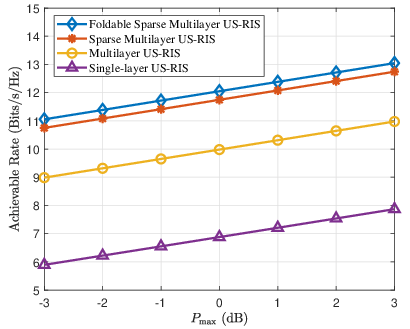}
\caption{Achievable rate versus transmit power.}
\label{fig:rate_192}
\end{figure}

\section{Numerical Results} \label{sec4}

In this section, we evaluate the performance of the proposed sparse and foldable sparse US-RIS architectures through extensive simulations. To ensure a fair comparison, we compare their achievable uplink rate with the conventional single-layer and multilayer US-RIS benchmarks to validate the effectiveness of combining element-wise and geometric sparsity.

\subsection{Simulation Setup}

In our simulation scenario, we consider an uplink communication setting in which the user and the BS are equipped with 2- and 8-element uniform linear arrays and are located at (0 m, 0 m, 0 m) and (0 m, 10 m, 0 m), respectively. The inter-layer spacing is set to $D=0.02$ m, the penetration loss factor is set to $\alpha=0.8$, the noise power is set to $\sigma^{2} = 10^{-6}$, and the signal frequency is set to $f=2.5$ GHz. In the simulation setup, perfect channel state information (CSI) is assumed to be available

For the evaluation, a total budget of $N_A = 256$ active elements is considered for the following architectures
\begin{itemize}
    \item \textbf{Single-Layer US-RIS:} A single layer with $256$ elements ($16 \times 16$ grid) located at (0 m, 0.02 m, 0 m).
    \item \textbf{Multilayer US-RIS:} Two layers, each with $128$ elements ($8 \times 16$ grid), placed at (0 m, 0.02 m, 0 m) and (0 m, 0.04 m, 0 m).
    \item \textbf{Sparse Multilayer US-RIS:} Three layers, each offering a grid of $128$ potential locations ($8 \times 16$). A total of 256 active elements are sparsely distributed across these 384 available grid points defined by $\mathbf{z}$.
    \item \textbf{Foldable Sparse Multilayer US-RIS:} Same configuration as the sparse multilayer case, with the added folding capability governed by $\mathbf{c}$ and $m=3$ feasible angles.
\end{itemize}

\subsection{Achievable Rate Analysis}

Fig.~\ref{fig:rate_192} presents the achievable rate versus the maximum transmit power, $P_{\text{max}}$, for different architectures. As expected, the rate increases monotonically with $P_{\text{max}}$. This shows that the introduction of sparsity provides a substantial performance boost, and the sparse multilayer US-RIS achieves a significantly higher rate than the conventional benchmarks. This improvement stems from distributing the same number of elements over a larger spatial aperture, resulting from an increase in the number of layers from two to three, which provides greater DoF for optimizing the beamforming. It is also clear that the best performance is achieved by the foldable sparse multilayer US-RIS, which leverages the additional geometric DoF from the folding configuration, $\mathbf{c}$, to further enhance the rate performance. 

\begin{figure}[t]
\centering

\begin{subfigure}[b]{0.5\columnwidth}
    \centering
    \includegraphics[width=\textwidth]{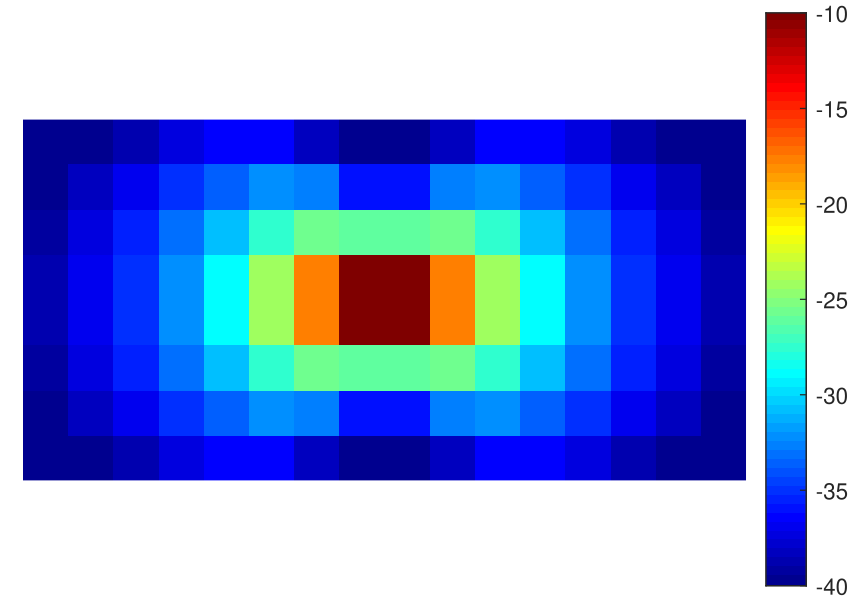}
    \caption{}
    \label{fig:ns_l1}
\end{subfigure}
\hfill
\begin{subfigure}[b]{0.5\columnwidth}
    \centering
    \includegraphics[width=\textwidth]{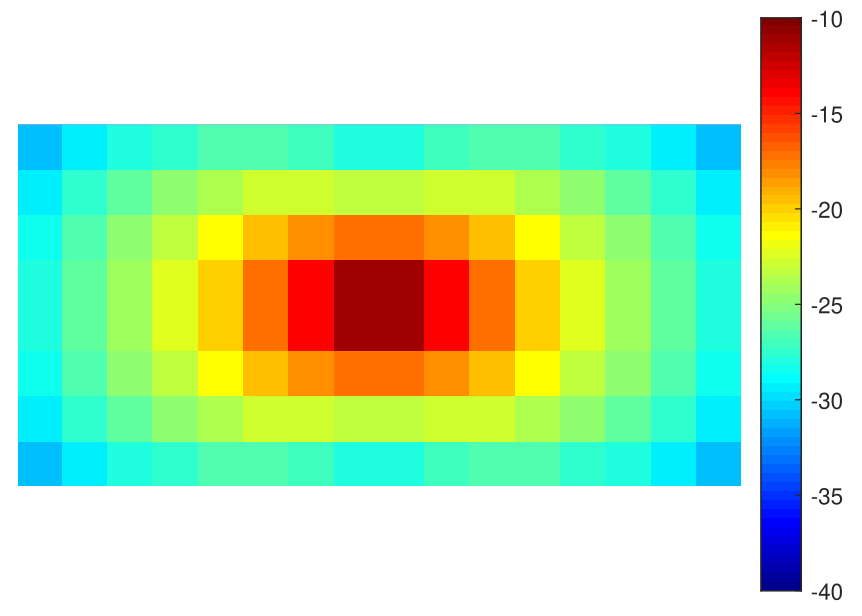}
    \caption{}
    \label{fig:ns_l2}
\end{subfigure}

\caption{Power distribution of non-sparse multilayer US-RIS: (a) Layer 1, (b) Layer 2.}
\label{fig:combined_power_analysis}
\end{figure}

\subsection{EAR Analysis}
\begin{figure}[t]
\centering

\begin{subfigure}[b]{0.5\columnwidth}
    \centering
    \includegraphics[width=\textwidth]{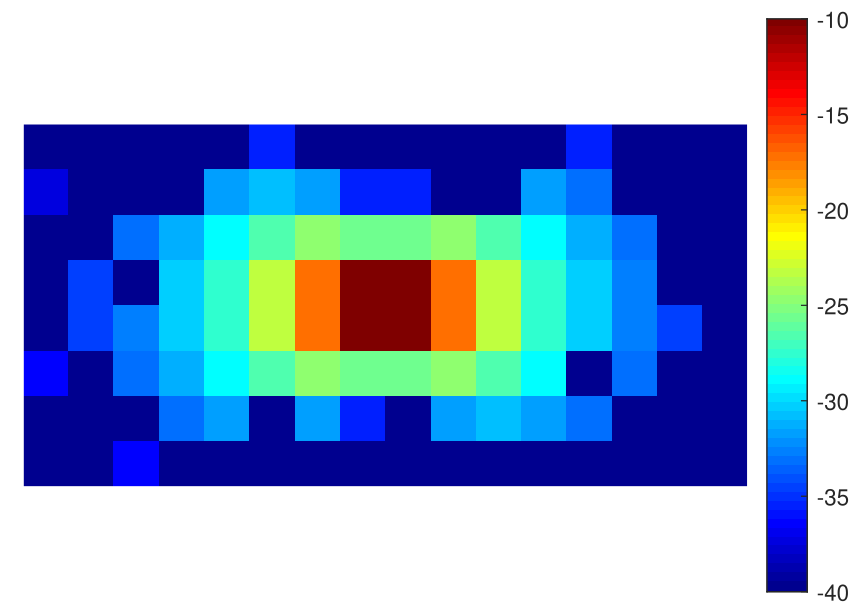}
    \caption{}
    \label{fig:fs_l1}
\end{subfigure}
\hfill
\begin{subfigure}[b]{0.5\columnwidth}
    \centering
    \includegraphics[width=\textwidth]{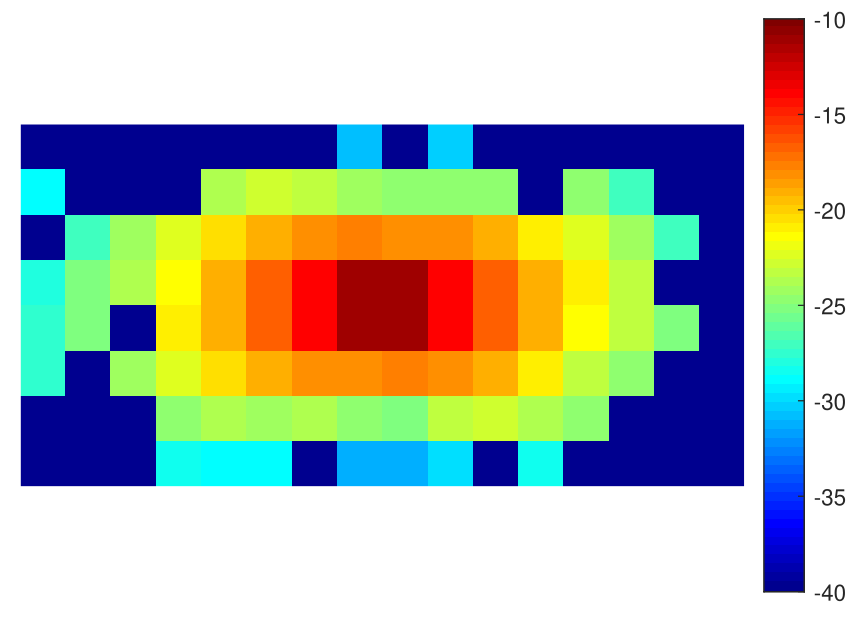}
    \caption{}
    \label{fig:fs_l2}
\end{subfigure}
\hfill
\begin{subfigure}[b]{0.5\columnwidth}
    \centering
    \includegraphics[width=\textwidth]{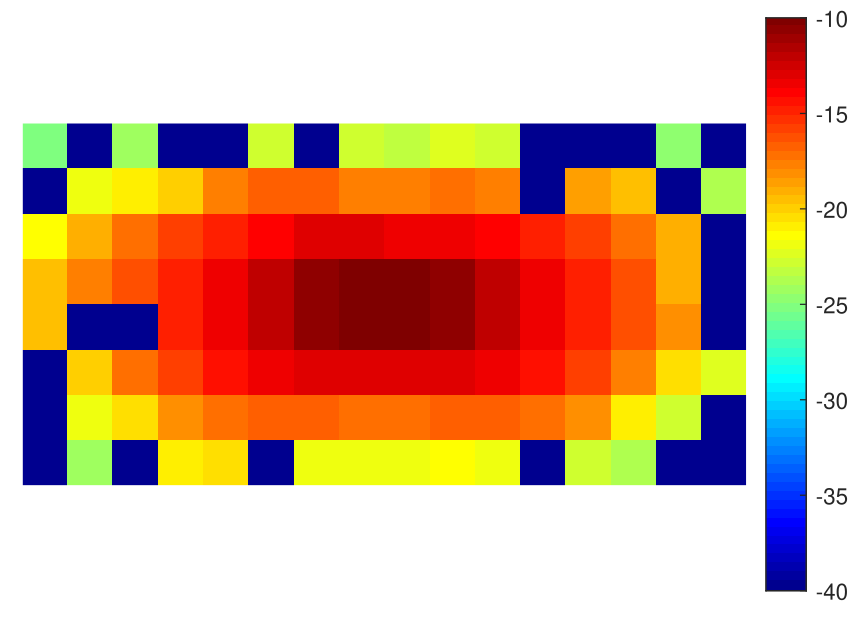}
    \caption{}
    \label{fig:fs_l3}
\end{subfigure}

\caption{Power distribution of optimized foldable sparse US-RIS: (a) Layer 1, (b) Layer 2, (c) Layer 3.}
\label{fig:combined_power_analysis}
\end{figure}

To understand how sparsity affects power distribution, we analyze the element activation ratio (EAR). For the sake of fair comparison, the EAR of the proposed foldable sparse US-RIS is compared to that of the conventional non-sparse US-RIS. Figs.~\ref{fig:ns_l1} and~\ref{fig:ns_l2} show that in the fully populated multilayer case, the power tends to be highly concentrated in the central elements, while in the foldable sparse configuration, the power is more effectively and broadly distributed across the active elements, as shown in Figs.~\ref{fig:fs_l1}-\ref{fig:fs_l3}. This improved hardware utilization is quantified by the EAR, which increases from 57.8\% in the two-layer non-sparse case to 72.77\% in the three-layer foldable sparse case. This improved EAR indicates that a larger proportion of active elements contribute significantly to the final beamforming gain.

\section{Conclusions} \label{sec5}

This letter presents a comprehensive framework for exploiting the benefits of sparse antenna arrays within the spatial constraints of compact user devices. In particular, two distinct sparsity paradigms for the US-RIS are proposed, namely element-wise sparsity and geometric sparsity. Furthermore, a novel form of geometric sparsity is introduced through a foldable RIS architecture. By optimizing the physical folding topology of the multilayer structure, the interlayer spacing is effectively utilized to create a larger effective aperture. Our simulation results validate that these proposed sparse architectures consistently outperform conventional single-layer and dense multilayer designs in terms of achievable rate. This work provides a practical solution for deploying high-performance large-scale arrays in future wireless user equipment.

\bibliographystyle{IEEEtran}

\bibliography{Ref}

\end{document}